# User Participation in an Academic Social Networking Service: A Survey of Open Group Users on Mendeley[1]

**Wei Jeng**

School of Information Sciences, University of Pittsburgh, 135 North Bellefield Avenue, Pittsburgh, PA, USA

Email: wej9@pitt.edu

**Daqing He**

School of Information Sciences, University of Pittsburgh, 135 North Bellefield Avenue, Pittsburgh, PA, USA

Email: dah44@pitt.edu

**Jiepu Jiang**

Center for Intelligent Information Retrieval, School of Computer Science, University of Massachusetts Amherst, MA, USA

Email: jiepu.jiang@gmail.com

**ABSTRACT**

Although there are a number of social networking services that specifically target scholars, little has been published about the actual practices and the usage of these so-called academic social networking services (ASNSs). To fill this gap, we explore the populations of academics who engage in social activities using an ASNS; as an indicator of further engagement, we also determine their various motivations for joining a group in ASNSs. Using groups and their members in Mendeley as the platform for our case study, we obtained 146 participant responses from our online survey about users' common activities, usage habits, and motivations for joining groups. Our results show that 1) participants did not engage with social-based features as frequently and actively as they engaged with research-based features, and 2) users who joined more groups seemed to have a stronger motivation to increase their professional visibility and to contribute the research articles they had read to the group reading list. Our results generate interesting insights into Mendeley's user populations, their activities, and their motivations relative to the social features of Mendeley. We also argue that further design of ASNSs is needed to take greater account of disciplinary differences in scholarly communication and to establish incentive mechanisms for encouraging user participation.

**Keywords**

Academic social networking, Mendeley, digital scholarship, online communities, research networks

**INTRODUCTION**

With the increasing popularity of the social web as well as the development of ever more powerful network technology, more and more scholars are joining online research communities. Taking advantage of the provided services, they often share academic resources, exchange opinions, follow each other's research, keep up with current research trends, and most importantly, build up their professional networks (Krause, 2012). Although non-academic social networking sites (SNSs) such as Facebook are much more popular, and scholars can communicate and collaborate with each other on SNSs, studies show that there are problems and limitations of using those sites to build academic users' professional networks (Gruzd, 2012). It is therefore advantageous for scholars to engage on social networking sites designed specifically for scholars – Academic Social Networking Services (ASNSs).

We use the term *academic social networking service* as a broad term that refers to an online service, tool, or platform that can help scholars to build their professional networks with other researchers and facilitate their various activities when conducting research. The term "ASNS" is related to multiple terms from different domains, such as *Networked Participatory Scholarship* (NPS) from the education field (Veletsianos & Kimmons, 2011) and *Research Networking* (RN) platforms or RN tools in health and biomedical-related fields (Schleyer et al., 2008; Weber et al., 2011). Some well-known examples of ASNSs include ResearchGate.net (http://www.researchgate.net/), Academia.edu (http://www.academia.edu/), Mendeley.com (http://www.mendeley.com/), and Zotero.org (http:// www. zotero.org/). ASNSs allow users to create profiles with academic properties, upload theirs publications, and create online groups (Oh & Jeng, 2011). Some ASNS, such as Mendeley and Zotero, even offer software applications, such as bibliographic tools to support scholars in managing their documents and citations.

Among the various social features of ASNS, online groups play an especially important role in connecting people with each other and with academic resources. From a user interface perspective, a group page in an ASNS can be viewed as a platform on which users can collaborate with their colleagues by sharing academic articles and conducting research discussions. For example, ResearchGate has a group function called "Project" that allows users to start a workspace with multiple "benches". Each bench can be used by the participants to exchange research data, articles, and ideas. Similar group functions for presenting and discussing trendy research topics are present in both Mendeley and Zotero as well.

Although there has been increasing interest in and understanding of the ways that non-academic SNSs (such as Facebook or Twitter) support scholars' research activities (Priem & Costello, 2010), few empirical studies have been carried out on scholars' usage of ASNSs. Of the studies that do exist, most focus more on the aspects that are related to predicted citation networks, scientometric studies, and new bibliometrics (a.k.a. altmetrics) (Li, Thelwall, & Giustini, 2011). There is little known about how academic users utilize the social features of an ASNS and how these social features can benefit users' online research experiences. To the best of our knowledge, there is a particular absence of studies that focus on scholars' activities in and usage of ASNSs' groups. As a result, we have an incomplete understanding of ASNSs' emerging services for scholars.

Our study, therefore, focuses on scholars' social activities and their usage in an ASNS and uses groups as the platform. Specifically, we want to answer the following research questions:

RQ1: Who are the users of an academic social networking service (ASNS) that supports open groups?

RQ2: In what ways, and how often, do such group participants use an ASNS?

RQ3: What motivates ASNS users to utilize social or research features on an ASNS?

We chose a cross-sectional survey as the research method to answer these research questions and developed a questionnaire with 30 questions. The questionnaire was distributed to 97 open groups in Mendeley, one of the most popular ASNSs. The process of data collection and the design details of the questionnaire will be covered in later sections.

We studied Mendeley group users with the goal of gaining insight into how they interacted with various features of the system. This knowledge enables researchers to know when and how an ASNS such as Mendeley can be appropriately included in their research activities. In addition, we want to support the further development of ASNS by answering questions related to the design of social features and incentive mechanisms.

**RELATED WORK**

**How Scholars Engage in General SNS**
Academic usage of social networking sites has recently started attracting more attention. Related works about how scholars engage in non-academic SNSs can be classified into two groups. Researchers in one group are interested in how a general SNS can support specific academic activities or functionalities. For example, Letierce et al. (2010) examined how Twitter can help people to follow the sessions and the topics covered at academic conferences. Researchers also pay attention to whether a general SNS can also be a platform for enhancing scholarly communication, such as the application of citations and scientific references on Twitter (Weller et al. , 2010; Weller & Puschmann, 2011).

Scholars in the second group focus more on to how well academic users adapt social networking sites in their daily activities. The studies in this group tend to depict the everyday experiences that scholars and scientists have on SNS sites. Through the examination of faculty members and scholars in higher education and research institutions, researchers learned that academic users found it difficult to establish boundaries between their personal and professional lives when using non-academic SNS. The main complaints included the loss of personal privacy and difficulties balancing work and life, which forced some academic users to create multiple accounts on an SNS (Gruzd, 2012). Academic social networking services emerged around 2008 in response to these new demands, offering scholars a new option for easing the tension between their personal lives and work lives. In the literature, researchers usually identified two groups of ASNSs, each of which originated from different online academic activities (Gruzd, 2012). Websites such as Academia.edu and ResearchGate started with supportive social-based features, whereas websites such as Mendeley and Zotero mainly emphasized research-oriented activities such as document and citation management (Zaugg, et al., 2011).

**Motivation for Joining Online Communities**
Studies have shown that personality, job characteristics, and prior life experience can all affect people's engagement in social activities (Kanfer et al., 2008; Latham & Pinder, 2005). Research focusing on online communities also paid more attention to the *functionalities* offered by an online groups or an SNS (Kietzmann et al., 2011). Thij (2007) classified the motivations people have for joining an online group as follows: social contact, information, financial or material benefits, support, interaction or discussion, and construction of self-identity. Butler and his colleagues (2002) categorized the perceived benefits of participating in online communities into four types: information benefit, visibility benefit, social benefit, and altruistic benefit. In order to capture possible factors that influences users' motivations in an ASNS, we measured both user characteristics and user motivations for joining an ASNS group in a response to related literature.

**Research on ASNS**
One major research focus for academic social networking services is related to new bibliometric or altmetrics. Though imperfect, citation-based bibliometric methods have been widely used to evaluate scholars for hiring, tenure, promotion, and other rewards and forms of recognition (Borgman, 2007). Altmetrics, on the other hand, examine the counts of users' bookmarks on articles in social reference websites or tags on social networking sites. The term "altmetrics" refers to " the creation and study of new metrics based on the social web for analyzing scholarship (Priem, Piwowar, & Hemminger, 2011)". Instances of data sources collected from social media include blogs (Shema, Bar-ilan, & Thelwall, 2013), BibSonomy bookmarks (Borrego & Fry, 2012), CiteULike bookmarks (Li, Thelwall, & Giustini, 2011), Mendeley reader counts (Bar-Ilan, Haustein, & Peters, 2012; Li et al., 2011), and Twitter mentions (Eysenbach, 2011). These studies found that most of the altmetrics were correlated with traditional publication indicators, which normally take years to accumulate.

While most of the studies mentioned above collected or analyzed a single case each, Thelwall et al. (2013) compared eleven social media websites and citations in the Web of Science database, suggesting that not all kinds of social media are suitable indicators for estimating scholarly impact. In particular, evidence was insufficient for LinkedIn, Pinterest, Social Q&A sites, and Reddit because of the lack of research components. Among all other websites with sufficient instances, mentions from Google+ were not significantly correlated with the Web of Science citations. This suggests that while data from *some* social media can be early an indicator of article impact and usefulness, not every source drawn from social media can successfully predict future citations.

In a study related to ASNS usage, Bullinger and her colleagues (2010) conducted in-depth interviews with the founders of ten websites including Academia.edu, Mendeley, and Research-Gate to develop a taxonomy for "social research networking service" (SRNS). In the same work, Bullinger classified SRNSs into four categories: 1) research directory sites, such as Academia.edu, that mainly focus on researchers' directory of contacts; 2) research awareness sites, such as ResearchGate and Mendeley, that enable self-promotion and profile management; 3) research management sites, such as Mendeley and CiteULike, that support users' research tasks and activities; and 4) research collaboration sites, such as CollabRx, that offer collaboration features. However, the boundaries between these categories are becoming increasingly blurred since most well-known ASNS usually offer all of the features described by the four categories (Oh & Jeng, 2011). Jeng et al. (2012) conducted a mixed-method study by qualitatively annotating owners' descriptions of Mendeley groups, and then applying statistical methods to examine the groups' member size and the collection size. Their findings suggest that a group is more likely to have more significant increases in membership and collection if the group description explicitly covers the scope and keywords through group owners' direct requirements.

There are also studies on the applicability of ASNSs as institutional repositories. Kelly et al. (2012) studied the presence of 20 UK universities on Academia.edu, LinkedIn, ResearcherID, and Google Scholar, and found that these research-oriented SNS can be help increase the visibility of researchers and their publications. Some scholars also noted that the reading list in ASNS groups can be seen as a virtual collection (Jeng, He, Jiang, & Zhang, 2012); further examinations are needed to verify the usefulness of this application.

Overall, despite the wide range of studies that have been carried out on ASNSs, studies on the usage of ASNS groups are nearly absent. In particular, the literature lacks conclusive insight into the reasons for scholars to use social features on ASNSs, their motivations for joining groups, as well as the perceived benefits of ASNS group usage. Therefore, the study presented in this paper helps to shift the focus back to the end users of ASNSs.

**RESEARCH SITE: MENDELEY**

Launched in 2008, Mendeley (http://www.mendeley.com/) is one of the most popular ASNSs and has more than two million users. Mendeley allows users to build their own digital research library by importing PDF files from their local devices. After an article is added to the library, its metadata is automatically extracted, including the title, authors, publication year, the name of the journal or proceedings, and so on. Users are able to open the imported PDF files, annotate them, take notes, and highlight text. Once an article is added to an individual's library, all users are able to search for the paper in the Mendeley catalog. As of December 2012, Mendeley's online catalog contained over 300 million research papers. As with common reference software (e.g. EndNotes), Mendeley offers capabilities to manage and generate citation data. Users can create bibliographies either from the papers they added or from the online catalog.

There are three common ways to use social features on Mendeley: maintain a profile, manage existing contacts, and make more connections. The "Profile" is a common feature for most social networking sites. Mendeley allows users to create a professional profile with research-oriented properties. For example, as shown in Figure 1, users can list their publications, research interests, advisees, awards, and grants on their own profile page.

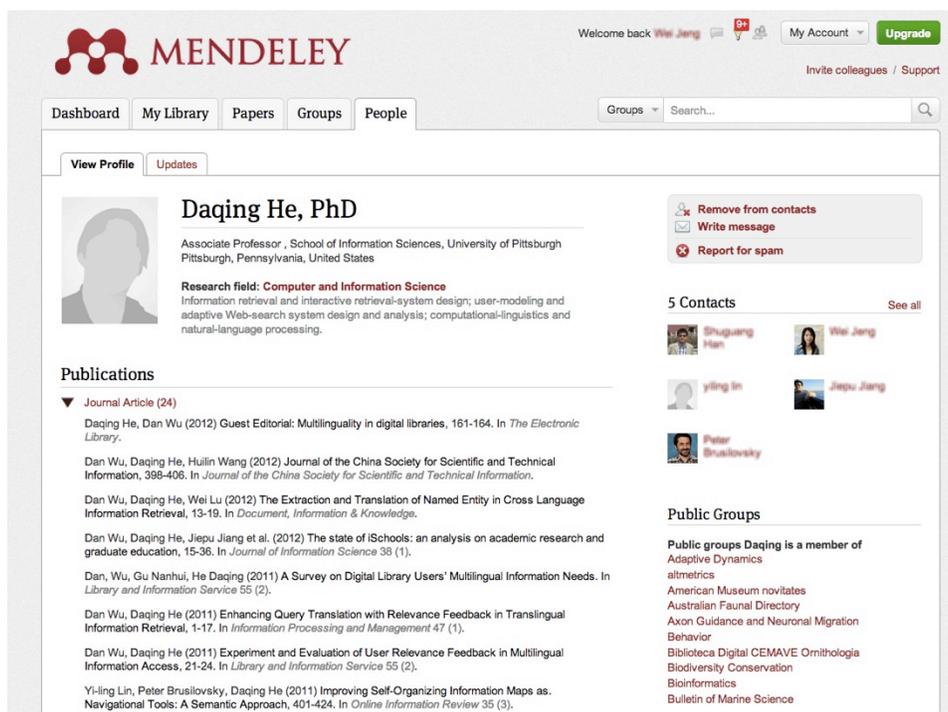

FIG 1. An example of a public user profile on Mendeley, a site similar to Linkedin. Mendeley allows users to fill in research-oriented properties such as publications, current affiliation, educational background, grants, research interests, and so on.

Mendeley also allows users to start groups to share what they are interested in and what they are reading about. There are two types of groups supported on the site: private groups that are only visible to the members and public groups that are publicly visible and can be searched in Mendeley's group list. The owners of public groups can decide whether membership is accessible to all users or if it has to be reviewed and approved. Thus, the public groups are either fully open or upon-approval. Users can freely join fully open groups, but need the owner's approval to join upon-approval groups.

Figure 2 shows an open group's "altmetrics" in Mendeley. Users can choose to be involved in a group either by joining the group as a member or following it as a follower. The only difference between these two roles is that members have the right to contribute articles. Both types of users are able to annotate an article as a "Like" and make other types of comments on the group's "Wall". The "About this group" panel shown in Figure 2 is a narrative introduction created by the group owner to inform new and potential group members about the intent of the group as well as the owner's expectations of the group. Other types of group content include the entire publication list, the member list, and the rules for interacting with other group members on the group's Wall. In this study, we are interested in scholars' social activities in ASNS. As shown in the aforementioned discussion of groups in Mendeley, ASNSs are potentially useful places to present and record scholars' social activities in an online environment. Given that Mendeley has nearly 77,000 public groups (May 2013), these groups can be important resources for studying whether and how scholars seek opportunities to form online collaboration. Based on the literature (Butler et al., 2002), we identified four types of potential motivations for scholars to utilize the group feature in Mendeley: to follow trending research topics, seek connections with like-minded people, gain a professional presence, and share their reading list with others. We will provide a detailed discussion of these motivations in the *Questionnaire Design* section.

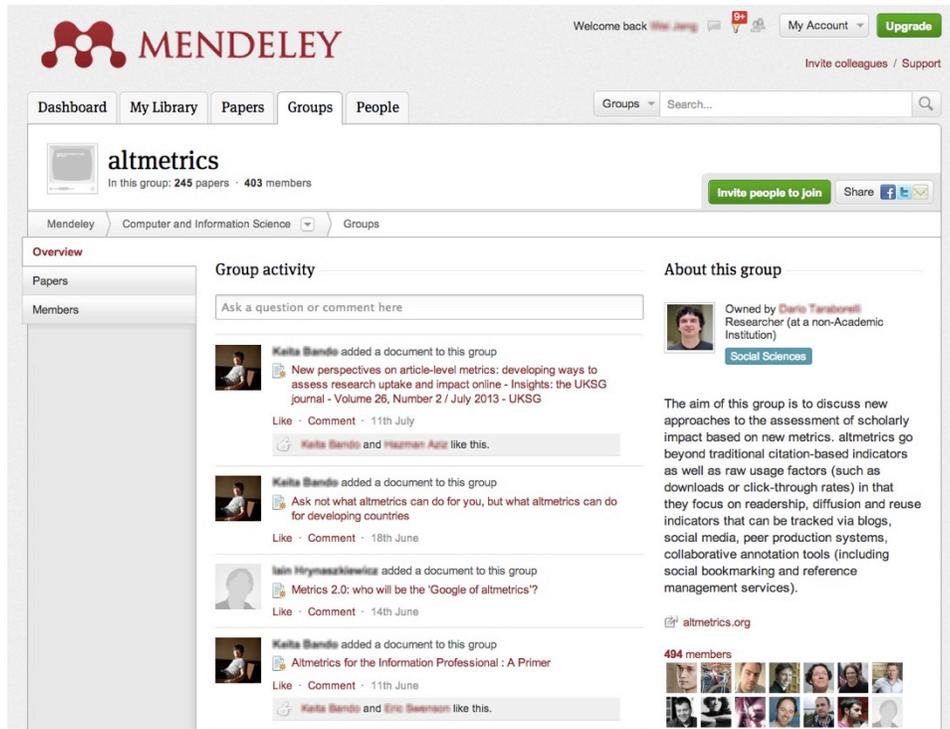

FIG 2. An example of an open group on Mendeley. Users are able to post content (e.g. an article), annotate an article as "Like" or make other types of comments.

**METHODOLOGY**

**Sampling and Piloting**

Before conducting the survey on Mendeley group users, we carried out a simple background study on group users in all Mendeley public groups (referred to as "background study" in Table 3). The goal of this background study was to better understand the group user population in Mendeley, which is important for answering RQ1 (who are the users of an ASNS). One of the authors wrote a script to automatically collect 54,703 members from all public groups (N= 34,838) in April 2012. Mendeley users who did not participate in any group were excluded. Of 54,703 users, only 7,366 (13.5%) provided detailed profile information (see Figure 1 for an example of such a profile). We then crawled these users' profiles to extract their job titles. The same author then manually sorted the various job titles into the following categories: faculty (e.g. "associate professor", "assistant professor"), postdoctoral fellow (e.g. "postdoc", "postdoctoral fellow"), doctoral student (e.g. "PhD student", "doctoral student"), graduate student, and librarian. We then executed a script to classify users' current positions from 7,366 profiles.

When performing the survey designed for this study, we adopted a representative sampling method to identify Mendeley's large group users. We specifically selected groups of more than 20 members based on the dataset used in a previous study (Jeng et al., 2012); therefore, we know that the Mendeley users we contacted are those engaged in Mendeley open groups. Our rationale for selecting large groups was that there would be a greater chance of social activities in large groups. The median number of group members in the sampled groups was 49, ranging from 20 to 284. Appendix 1 provides a list of sampled groups (N=97).

Our data collection method was a questionnaire-based survey. We ran an initial pilot questionnaire on a convenience sample of 13 doctoral students at two research universities in Pittsburgh, USA. This helped us to revise the design of the questionnaire, and the modified survey was conducted over a two-month period from October to December 2012 using an online questionnaire software program (Qualtrics). One of authors manually posted questionnaire links to 97 open groups on Mendeley. Three weeks after the initial posting, the author posted a follow-up message beneath the post to encourage more participation. No incentives were provided.

**Questionnaire Design**

The final questionnaire contained one open-ended question and 29 Likert-scale questions (see Appendix 2). Five of the 29 questions were developed to collect users' basic demographic information for RQ1. We then designed 15 items to answer RQ2, which included questions about the participants' ways and degrees of utilizing Mendeley and its social features. Nine questions and the one open-ended item were used to ask academic users about their motivations for joining a group. The answers to the rating questions are marked as 5-point Likert scales with "1" being the lowest degree and "5" being the highest degree. Each group's ID was embedded in the questionnaire in order to monitor the response rate of sampled groups.

The first part of the questionnaire contained basic demographic questions.

*Basic Information*. For RQ1, we collected four types of background information: the participants' disciplines, job positions, gender, and age. It is important to note that when a user registers in Mendeley, the individual has to choose one scholarly discipline, such as education or biology, from 25 predefined categories to indicate his or her domain. The position information reflects the users' academic jobs (e.g. faculty, doctoral student, librarian, and so forth).

We collected these four types of information to see if they are potential factors that associated with the motivations for joining groups, as the literature suggests (Latham & Pinder, 2005).

The questions related to RQ2 (how and how often do group participants use an ASNS) are further divided into the subcategories below.

*The Extent of Use*. These questions aim to determine the extent of participants' account activities on Mendeley, including how frequently users visit their accounts, update their profiles, and check their news feeds (i.e. item S.1.1 – S.1.5 in Appendix 2).

*Common Ways to Use*. The questions in this subcategory concern the six most common ways of using Mendeley: as a document management tool, a reference manager, a scholarly search engine, an online portfolio, a friend management tool, and a socialization tool (i.e. item C.1 – C. 6 in Appendix 2). We posed these questions to explore the various ways that scholars may use an ASNS site and whether these diverse usages would influence the likelihood that scholars would embrace social features on ASNSs. We specifically asked participants if they used Mendeley to interact with their existing friends or to connect with new online contacts only. Our rationale, based on a previous study (Nelson Laird et al., 2008), was to differentiate between users' offline and online networks since the nature of these two networks can be different.

*The Extent of Group Use*. Items in the "Engagement of Group Use" in Appendix 2 show questions related to users' actual engagement in Mendeley groups, such as the number of groups they created, joined, and followed (i.e. Item G.1.1- G.1.4).

To address RQ3 (what motivates ASNS users to utilize an ASNS), we asked the participants to talk about their motivations for joining Mendeley groups. In keeping with previous literature (Bateman, Gray, & Butler, 2006; Butler et al., 2002; Kietzmann et al., 2011; Thij, 2007), we identified four types of motivations. We used nine indirect questions to detect the types of reasons for joining a group. All items were followed by a verbatim question: "Based on the overall impressions of your past activities, you usually join a Mendeley group for __". To help readers better understand how we constructed the items about these four motivations, we will present them in detail below.

*Motivation 1: Seeking Information*. Researchers found that users' informational needs can motivate them to participate in online communities (Lampe et al. , 2010). Prior studies also suggested that people often participate in an online community in order to acquire user-generated content or to solve their own information problems (Porter, 2004). For academic users, information seeking and acquisition behaviors such as the literature search, bibliographic search, and the factual information validation are very common in their online academic activities (Meho & Tibbo, 2003). To capture the different information-seeking behaviors in both informational and navigational tasks, we designed three corresponding questions to define users' motivation for joining open groups (i.e. Item M1.1 – M1.3 in Appendix 2).

*Motivation 2: Seeking Networks and Connections.* People also join communities to fulfill social needs, such as connection forming or finding like-minded people (Cummings, Sproull, & Kiesler, 2002; Latham & Pinder, 2005). An ASNS group page is a platform that can bring people with similar research interests together and even form a community. Users are able to build their social ties and personal networks through the group. For example, following a familiar Facebook interface, Mendeley group pages show a list of group members and allow users to connect to others in the same group. We propose that an ASNS group exists to satisfy users' needs of networking and building connections. Four items followed by the same verbatim question (noted in the previous paragraph) were used to identify group members' intentions of building networks (i.e. item M2.1 – M2.4).

*Motivation 3: Gaining Professional Visibility.* Gaining personal presence is also an important motivation in a social networking environment. Why would scholars need to establish a presence? Leahey (2007) indicated that the increase in professional visibility for scholars has a positive and significant effect on rewards (e.g. salary, reputation, and positions). Beyond the benefits gained by scholarly publishing, we propose that users are also able to enhance their professional presence by being present in discussions and engaging in other group activities. We used the two items M3.1 and M3.2 in Appendix 2 to measure the intention of gaining visibility for people to join groups on Mendeley.

*Motivation 4: Showing Altruistic Actions.* An altruistic action is a recognizable phenomenon in the real world as well as in the virtual world (Lampel & Bhalla, 2007). In the real world of scholarly community, people are motivated to help others (Kogan, 2000). Given that users join many online communities voluntarily, we intended to determine if online scholars are also motivated to participate in online group activities in order to make an altruistic contribution. Specifically, we asked group members if they joined a group in order to "contribute to the reading list" (i.e. item M4 in Appendix 2).

**RESULT**

**The Participants**

In total, we received 188 responses via the questionnaire, but only 146 users completed the entire questionnaire. Therefore, the analyses in this section are based on these 146 complete responses. The average age of the participants was 35.04 years (SD=10.81), and 64% of them were male (N=94).

We obtained responses from users in 20 disciplines in Mendeley. The top three disciplines represented were computer and information science (N=43), biological science (N=24), and social science (N=17). The results from the top 3 disciplines were generally consistent with the distribution in Oh and Jeng's study (2011), where the data were collected from the official API data in Mendeley. In order to apply valid statistical methods when comparing the differences among 20 disciplines, we merged several disciplines into a larger "discipline group" (see Table 1).

**Table 1. Service Adaptation by Discipline Groups**

| Discipline Groups | Adaptation | | | | Total | |
|---|---|---|---|---|---|---|
| | < 6 months | 6m-1yr | 1-3 year | 3+ yr | N | % |
| 1. Computer and Information | 9 | 10 | 21 | 2 | 43 | 29.5 |
| 2. Social Sci, Edu, Law, and Psy | 11 | 11 | 13 | 1 | 36 | 24.7 |
| 3. Biomedicine | 5 | 5 | 18 | 4 | 32 | 21.9 |
| 4. Other STEM disciplines[b] | 7 | 4 | 8 | 1 | 20 | 13.7 |
| 5. Business, Management, and Economic groups | 3 | 0 | 7 | 0 | 10 | 6.8 |
| 6. Art and Design disciplines[c] | 3 | 1 | 1 | 0 | 5 | 3.4 |
| Total | 38 | 31 | 68 | 8 | 145[a] | 100 |

Note: a. One user in Computer and Info. Science answered "unsure".
b. Other STEM Disciplines in this study included astronomy, chemistry, earth sciences, engineering, environmental sciences, materials science, mathematics, and physics.
c. The uncategorized disciplines included arts and literature (N=2), design (N=1), and humanities (N=2).

In Table 1, we also drilled down into the distribution of disciplines in terms of how long users have been using Mendeley. The discipline grouping method we conducted referred to the taxonomy of Biglan Categories (as cited in Nelson Laird et al., 2008). Data shows that the majority of participants (N=22, 68.8%) in biomedicine groups started their Mendeley account more than one year earlier, while more users (N=22, 61.1%) in the social sciences (including social science, education, law and psychology) group started to use Mendeley within a year. It is worth noting that the discipline group of humanities, literature, philosophy, and design was relatively absent from our sample when compared to other discipline groups.

Table 2 shows the users' position distribution in the current survey in comparison to another sample (N=7366) that we collected from the Mendeley webpage in April 2012. Among our sampled users, the majority of the participants in the current study were doctoral students (N=63, 47%). Of the remaining participants, 16 were master's students, 14 were faculty members, 13 were postdoctoral researchers, 9 were librarians, and 9 were undergraduate students. Even though the rank distribution varied slightly in the two samples, we still found that Mendeley users were in academia or involved in research-oriented jobs (e.g. doctoral student, faculty member, and postdoctoral position). The rest of the participants held positions in fields related to the higher education environment and academia, such as librarians and industrial researchers.

**Table 2. Sample Distribution of Job Positions**

| Job Title | Current Study: N | % | Background Study[a] N | % |
|---|---|---|---|---|
| B.2.1. Doctoral student | 66 | 45.2 | 1968 | 26.7 |
| B.2.2. Graduate student | 21 | 14.4 | 270 | 3.7 |
| B.2.3. Faculty member | 15 | 10.3 | 1776 | 24.1 |
| B.2.4. Postdoctoral researcher | 14 | 9.6 | 592 | 8.0 |
| B.2.5. Industrial researcher | 2 | 1.4 | 206 | 2.8 |
| B.2.6. Undergraduate student | 10 | 6.8 | -- | -- |
| B.2.7. Other student [b] | -- | -- | 198 | 2.7 |
| B.2.8. Librarian | 9 | 6.2 | 168 | 2.3 |
| B.2.9. Other research professional | 9 | 6.2 | 2188[c] | 29.7 |
| Total | 134 | 100 | 7366 | 100 |

Note: a. Among 54,703 group users, only 13.5% (N=7,366) completed detailed profile information.
b. Students who cannot be categorized into doctoral and graduate student
c. Position directly denoted that research-oriented but cannot be categorized.

**How Scholars Use Mendeley**

The results in Table 3 show that 53% of respondents visited their accounts on a weekly basis, while 36% of them accessed the site at least once per month. However, more than half (53%) of the participants reported that they were checking the news feeds only on a monthly basis, not as frequently as visiting their accounts. The user profile page on Mendeley is similar to a page on LinkedIn or other professional site and is basically a static element similar to a portfolio or a web CV. The majority of the users (57%) in our study updated their Mendeley profiles several times a year or even less, which was consistent with the previous survey study (Skeels & Grudin, 2009).

**Table 3. Frequency of Account Activities**

| Items | Weekly | Monthly[b] | Annually | Never |
|---|---|---|---|---|
| S.1.1 Account visiting | **53.4%** | 35.6% | 10.3% | .7% |
| S.1.2 News feed checking | 14.4% | **53.4%** | 20.5% | 11.6% |
| S.1.3 Profile updating | 2.7% | 24.0% | **56.8%** | 16.4% |

Note: a. All items in this table followed a "how-often" question, N=146. (1= never, 2=several times a year, 3= once a month, 4= 2-3 times a month, 5= once a week or more).
b. The percentage of answers that were the sum of 3 and 4.

In Table 4, we asked participants to report on their usual ways of using Mendeley. Ranked by the Mode of the participants' responses, participants primarily used Mendeley as a document management tool and as citation management software, respectively. Only 13% (including degree of "4" and "5") of the respondents used their profiles as an online portfolio or a web CV. The portion of those using Mendeley as a social networking site was relatively low: Only 11% of respondents used Mendeley to manage their existing academic friends in degree "4" and "5", and 9% used it to expand their professional networks. These results indicate that most of our participants use Mendeley for its research features.

**Table 4. Ways to Use Mendeley**

| Theme | Items | 1(low) | 2 | 3 | 4 | 5 (high) | M | SD |
|---|---|---|---|---|---|---|---|---|
| Research features | C.1 As doc management | 3.4% | 10.3% | 15.1% | 28.1% | **43.2%** | 3.88 | 1.224 |
|  | C.2 As a reference manager | 8.9% | 9.6% | 21.9% | 23.3% | **36.3%** | 3.66 | 1.301 |
|  | C.3 Scholar search engine | 6.8% | 20.5% | **38.4%** | 19.9% | 14.4% | 3.03 | 1.179 |
| Social features | C.4 As an online portfolio | 22.6% | **39.0%** | 25.3% | 10.3% | 2.7% | 2.26 | 1.016 |
|  | C.5 Manage existing friends | **39.0%** | 32.2% | 17.8% | 6.2% | 4.8% | 2.01 | 1.075 |
|  | C.6 Make more connections | **37.7%** | 31.5% | 21.9% | 6.2% | 2.7% | 1.97 | 1.011 |

Note: All items were followed by the question, "To what degree do you use the following features?", N=146, and the participants responded using a 5-point Likert Scale.

Table 5 illustrates the numbers of groups that users have created, joined, or followed as well as the contacts they had previously added to their list. The majority of the participants had never started a group, and none had been the owner of a private group (N=90, 62%) or a public one (N=101, 69%). 37% and 26% of the participants had created 1-5 private groups and public groups, respectively. 69% of the respondents (N=100) reported that they joined 1-5 groups (both private and public), while 59% of the participants (N=86) had followed 1-5 groups that were created by others. These results suggest that the majority of the participants have noticed the group function and have been part of groups that were created by others, but not many had experience creating a group on the research site.

**Table 5. Group Use and Numbers of Contacts**

| Items[a] | 10+ | 6-10 | 1-5 | None |
|---|---|---|---|---|
| G.1.1 Create N Private groups | .7% | .7% | 37.0% | **61.6%** |
| G.1.2 Create N Public groups | .7% | 4.1% | 26.0% | **69.2%** |
| G.1.3 Join N groups | 6.2% | 16.4% | **68.5%** | 8.9% |
| G.1.4 Follow N groups | 3.4% | 15.1% | **58.9%** | 22.6% |
| S.1.5 Have N contacts | 26.7% | 23.3% | **38.4%** | 11.6% |

Note: All items in this table followed a "how-many" question, N=146. (1= none, 2=1-5, 3= 6-10, 4= 11-20, 5- 20+, 6=100+).

**Why Users Join Mendeley Groups**

After removing the participants who never joined a group on Mendeley, the remaining sample was down to 130. Table 6 reports the items of users' motivation in terms of joining a Mendeley group created by others. As shown in the table, the top 2 motivations for joining a group were keeping up with a user's research domain and following topics that the community is paying attention to. The motivations of expanding current social networks and keeping in touch with current contacts received a lesser degree of agreement.

**Table 6. Motivations for Joining Groups (N=130)**

| Items[a] | | Cronbach's α | 1(low) | 2 | 3 | 4 | 5(high) | M | SD |
|---|---|---|---|---|---|---|---|---|---|
| M.1.1 | Keep up with a user's research domain | .702 | 1.5% | 3.1% | 8.5% | 37.7% | **49.2%** | 4.30 | .868 |
| M.1.2 | Get research-related questions answered | | 7.7% | 11.5% | 29.2% | **35.4%** | 16.2% | 3.41 | 1.125 |
| M.1.3 | Follow topics that community is paying attention to | | 5.4% | 3.1% | 12.3% | **46.2%** | 33.1% | 3.98 | 1.034 |
| M.2.1 | Connect with people who have similar research interests | .875 | 3.1% | 6.2% | 21.5% | **35.4%** | 33.8% | 3.91 | 1.038 |
| M.2.2 | Expand current social network | | 5.4% | 23.1% | 28.5% | **29.2%** | 13.8% | 3.23 | 1.117 |
| M.2.3 | Meet more academic people | | 6.9% | 14.6% | 23.1% | **35.4%** | 20.0% | 3.27 | 1.169 |
| M.2.4 | Keep in touch with people one already knows | | 13.8% | 16.9% | **28.5%** | **28.5%** | 12.3% | 3.08 | 1.227 |
| M.3.1 | Gain professional visibility | .674 | 6.2% | 12.3% | 26.2% | **38.5%** | 16.9% | 3.48 | 1.101 |
| M.3.2 | Be present in current discussions | | 8.5% | 10.8% | **36.9%** | 33.1% | 10.8% | 3.27 | 1.070 |
| M.4 | Contribute to the reading list | -- | 6.2% | 7.7% | 23.1% | **43.8%** | 19.2% | 3.62 | 1.073 |

Note: All items in this table followed a "why join" question, N=130. (1= strongly disagree, 5= strongly agree).

A Cronbach's α of the construct was used to obtain the reliability of each type of motivation: information (α=.702), networking (α=.875), visibility (α=.674), and altruistic (only one item). The Box-and-Whisker Plot in Figure 3 represents the median, lower quartile, higher quartile, range, and the outliers of all types of motivations. Generally speaking, users had stronger intentions to seek information and to perform altruistic behaviors (Mdn = 4.00) than to seek visibility and engage in networking (Mdn=3.50) when considering whether or not to join Mendeley groups.

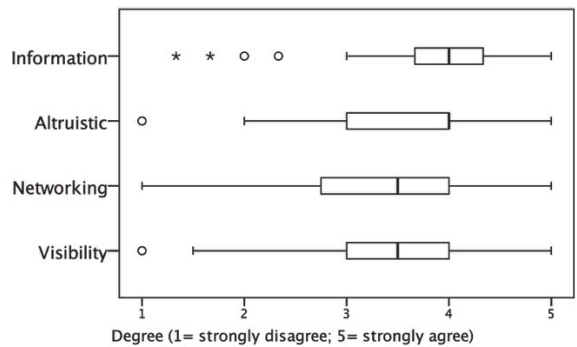

FIG 3. The Box-and-Whisker Plot represents the motivation that users perceived when joining a group. The information-seeking and altruistic motivations have a higher median than the networking and visibility motivations.

As shown in Figure 4, by (or when) comparing the differences in users' job characteristics, we observed that faculty members seemed to have the lowest networking motivation (N=14, Mdn= 3.00), and the industrial researcher and other professionals in the same category had the highest value (N=11, Mdn= 3.62). Undergraduate students (N=9, Mdn= 3.00) and postdoctoral researchers (N=13, Mdn=3.00) had lower visibility motivations than any other user group. However, a Kruskal-Wallis test showed no statistical difference across four categories of motivations in terms of job characteristics.

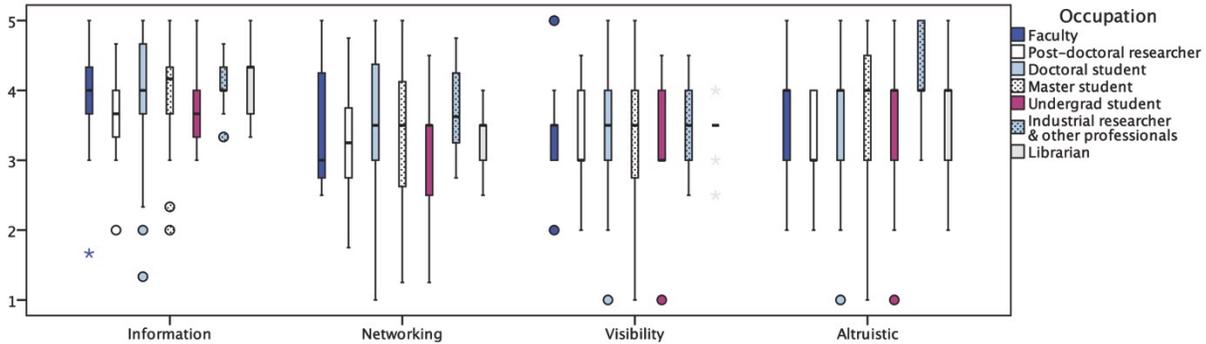

FIG4. Motivations for group joining by users' job characteristics (occupation).

Figure 5 presents how differently each gender perceived the motivations for joining a group. A Mann-Whitney U test found that female users (N=37) had significantly stronger motivations compared to males on information, U=1195, p<. 05, and on altruistic, U=1114, p<. 01. Figure 6 illustrates the motivations across three disciplinary areas. A Kruskal-Wallis test showed no statistical difference across disciplinary groups.

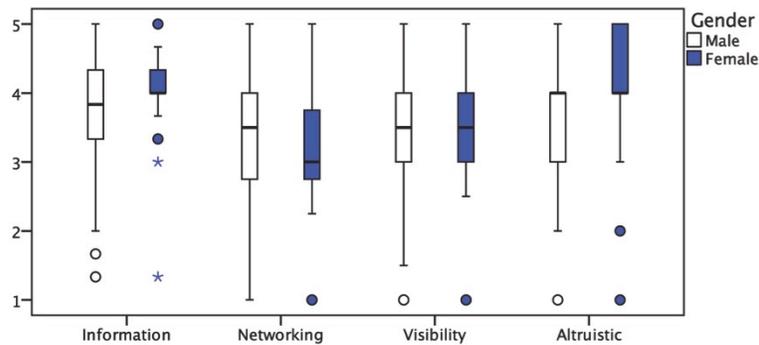

FIG 5. Motivations for group joining by users' gender

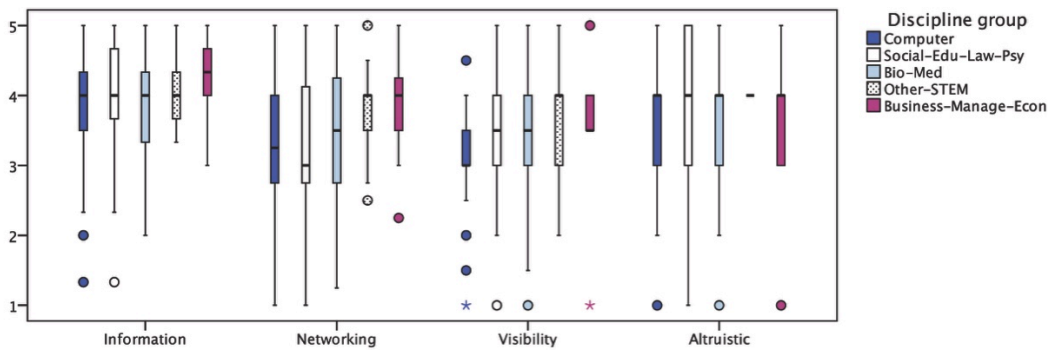

FIG 6. Motivations for group joining by users' discipline areas (discipline).

In Table 7, we extracted 13 items that we obtained with regards to RQ1. We selected these items as factors and explored whether they had any influence on the outcome of joining motives. To our surprise, there was no significant difference of any kind in terms of account accessing (S.1.1) and numbers of contacts (S.1.5). Instead, users who updated their profiles monthly had a significantly higher level of information, networking, and visibility motivation than people who updated their profiles less often. It is worth knowing that users who checked their news feeds on a weekly basis had the highest outcome of all kinds of motivation at the .05 level. Surprisingly, users who had joined six or more groups (G.1.3) tended to have higher intentions of altruistic behavior than people who were in five or less groups. We obtained a similar pattern regarding being followers (G.1.4). On the other hand, no matter how many group users joined or followed, we did not observe a significant difference in terms of the change in their networking intention. These results suggest that people are most motivated by visibility and altruism when considering whether to join or follow more groups (e.g. 6 or more). We also found that the most common ways of using Mendeley (such as a document management or citation management tool) failed to associate with the outcome of any group-joining motives, while other features that users preferred (C.3-C.6) were correlated to all types of motivations except altruistic. These results suggest that even if users frequently and regularly engaged in research-based activities on Mendeley, it would not make any difference in terms of their intentions of joining groups.

**Table 7. Factors Comparison**

| Perceived factors (N=119[c]) | Information | | Networking | | Visibility | | Altruistic | |
|---|---|---|---|---|---|---|---|---|
| | $\chi^2$ | p | $\chi^2$ | p | $\chi^2$ | p | $\chi^2$ | p |
| S.1.1 Account accessing | .889 | .641 | .403 | .817 | .228 | .892 | .299 | .861 |
| S.1.3 Profile updating | 12.859 | .002** | 19.412 | .000** | 15.370 | .000** | 2.658 | .265 |
| S.1.2 News feed checking | 16.562 | .000** | 11.213 | .004** | 11.547 | .003** | 7.958 | .019* |
| G.1.3 Being in N group as a member[d] | 3.247 | .072 | 1.763 | .184 | 3.222 | .073 | 6.535 | .011* |
| G.1.4 Being in N group as a follower | 8.905 | .012* | 2.934 | .231 | 8.687 | .013* | 9.852 | .007** |
| S.1.5 The no. of contacts | .291 | .865 | 1.936 | .380 | .321 | .852 | .602 | .740 |
| C.1 Document management | .669 | .716 | .570 | .752 | 2.616 | .270 | 2.935 | .231 |
| C.2 Citation management | 1.440 | .487 | .224 | .894 | .141 | .932 | .211 | .900 |
| C.3 Search papers | 18.286 | .000** | 11.826 | .003** | 17.501 | .000** | 7.813 | .020* |
| C.4 Managing professional profiles | 11.073 | .004** | 10.619 | .005** | 9.641 | .008** | 6.599 | .037* |
| C.5 Keeping existing friends | 16.568 | .000** | 33.238 | .000** | 17.694 | .000** | 5.445 | .066 |
| C.6 Exploring network | 15.823 | .000** | 26.088 | .000** | 21.147 | .000** | 7.574 | .023* |

Note: *: p<0.05; **: p<0.01.
a. N= 130, All data were ranked before a Kruskal-Wallis H test ($\chi^2$).
b. All factors in Table 7 were recoded into three conditions (the lowest degree, medium degree, and the highest degree) to determine if different conditions would associate with the outcome of group joining motives differently, df=2. Item ID in Table 7 can be mapping to the items in Table 1.
c. S.1.1-S.1.3 were recoded into seasonally or less, monthly, and weekly. S.1.5, G.1.3 and G.1.4 were recoded into none, 1-5, 6 & above. Items C.1-C.6 were also recoded into lower degree, medium, and higher degree.

## DISCUSSION & IMPLICATIONS

### Insights Based on the Findings

*Mendeley: A Platform for Higher Education Users*
Our findings confirmed that the majority of Mendeley users were from the higher education environment. More specifically, as Table 3 shows, users in three of the top four categories in both this study and our background study were identified as junior researchers (i.e., doctoral students, post-doctoral fellows, and graduate students). Thus, for those researchers who would like to study junior scholars' information behaviors or run a survey on a wide range of online scholars, we believe that an ASNS such as Mendeley are the right platforms to use to reach those types of participants. We do acknowledge that our samples have more doctoral students but less faculty members than the proportion obtained in one of our background studies on Mendeley users. This inconsistency might influence the generalizability of our findings. On the other hand, this distribution might reflect the actual practice of engaging open groups instead of all registered users on Mendeley, since faculty members might lack the time to actively engage in large open groups. We also would like to warn that there is a bias in our study in terms of a discipline distribution in Mendeley. We will discuss the distribution in the next sub-section.

*Discipline Distribution and Development in Mendeley*
Our results show that the discipline development in Mendeley is not evenly distributed (Table 2). Early users in Mendeley groups mostly came from the fields of computer & information science and biomedicine, whereas more recent users are mostly from the fields of social science, education and psychology. Additionally, the number of groups indicates that disciplines such as computer and information science, social science, education, psychology, and biomedicine are popular in Mendeley groups. At the same time, we do not see many group users from the humanities and other related fields. The community preference is one possible explanation for this disparity. For example, Zotero might be more popular in the historian and the archivist community (Cohen, 2007; Rogers, 2008), which might affect the number of historians who join Mendeley. Another possible reason can be derived from the work-related activities and the "information styles" of a specific discipline (Palmer, 1991). For example, computer science users might have more opportunities to engage in Internet activities while they work, whereas humanity scholars might have different work habits and ways of acquiring and sharing information relative to their daily routine, such as literature comparing, interpreting, and critiquing (Case, 2006). Lastly, we also assume that the practice of choosing large groups may be biased towards certain communities such as biological scientists, who have adopted community data-sharing practices such as the GENBANK (www.ncbi.nlm.nih.gov/genbank/) in the National Center for Biotechnology Information (NCBI). An important future research direction is to examine whether this unevenness is caused by community preference, community data-sharing practices, disciplinary characteristics, or a lack of adequate support in ASNS for certain disciplines. We believe this insight will help to create a better understanding of the academic ecology of ASNS.

*Academic Social Networking: "Academic" but not "Social"?*

The results of our study suggest that the participants do not engage with social networking features as frequently and actively as with research-based activities. In other words, users of Mendeley seem to mainly concentrate on the utilities directly related to their research work, while mostly ignoring its social features, such as "friend making".

We do not know the exact reasons for this uneven usage of Mendeley's research and social features. The lack of socialization among Mendeley users can have several possible explanations. One of these reasons may not be related to ASNS, but may come from the nature of shared expertise. Previous work pointed out that users sharing expertise online might be challenged by difficulties conveying their knowledge or understanding to others. They may also be concerned about sharing too much critical information and may feel that it is too time-consuming to explicitly articulate their knowledge to others. (Hinds & Pfeffer, 2003). Given that ASNS serve as a new and innovative channel for scholars that did not exist before, some of these challenges might eventually be overcome. Therefore, it deserves further study on whether an ASNS still has room for improvement in terms of supporting knowledge exchange. Another possible reason may originate from the first-mover advantage. Compared to leading non-academic social networking services, ASNSs came into being only relatively recently and are much smaller in scale. That is, academic users may already be established and feel more comfortable in conducting their social activities on Facebook or using LinkedIn to manage their professional social life; there is no immediate and compelling reason for them to replicate their same social networking activities in an ASNS such as Mendeley. If this is true, developers of ASNS might want to reconsider whether to offer social features at all and may need to work on making them innovative enough to compete with Facebook and LinkedIn. No matter the reason, we believe that our findings serve as a warning for ASNS developers to think carefully about simply adopting "Facebook-like" or "LinkedIn-like" social elements when designing an ASNS platform for academic users.

*Join a Group? Show Me the Incentive.*

The anticipated incentives for academic users' scholarly activities include material reward, recognition for tenure or promotion, and academic outcomes (which can also be called extrinsic motivations). However, academic users can also be motivated by self-efficacy beliefs and altruistic behaviors. Our results from exploring academic users' utilization of Mendeley groups showed that altruistic motivation was one of the most critical reasons associated with their group engagement, yet none of current features of Mendeley reward scholars for their altruistic activities. We believe that encouraging or facilitating altruistic behaviors in Mendeley is necessary because volunteering always plays an important role in online communities. In this sense, designing multiple incentive mechanisms is important for all ASNSs too. Even though ResearchGate launched the RG Score (https://www.researchgate.net/publicprofile.RGScoreFAQ.html) as an alternative metric to represent users' scientific reputation, the intrinsic motives of academic users are still overlooked. Possible incentive mechanisms that could work both in Mendeley and in general ASNSs could include the providing of affective feedback by group owners or members, such as a warm greeting or a simple "Like", to users who answer others' questions or contribute to a list of readings. ASNSs could also establish some form of elaborate level-based honor system, such as the user levels in Yahoo! Answers. Social gaming features like the badging system in Foursquare might be another interesting approach to encourage interactions.

**Limitations**

In order to ensure the response rate and find more representative users who tend to engage in social networking features on an academic social networking service, we sampled only open and large groups with many members on Mendeley instead of a random sample. Therefore, the discipline and position distribution of the participants in our survey may not be totally consistent with the entire population of ASNS users. This bias might affect the current study's outcome when we compared different motivations by the regrouped discipline area (Figure 5). It may also be biased towards users who are the use group of social feature and highly engaged in group activities. In order to understand the whole picture of online research groups, studying upon-approval and private groups is essential; if researchers would like to investigate the wider landscape of ASNS users, larger-scaled and random sampling approaches are needed.

**CONCLUSION**

In this study, we examined the users' practices and motivations for joining an online research group in an academic social networking service (ASNS) — Mendeley. To the best of our knowledge, this is the first study that investigates users' attitudes and motivations for joining groups in an ASNS.

By analyzing the data directly provided by the users themselves, we confirmed that Mendeley users were primarily academics, especially junior researchers. Our results also show that users of Mendeley do not explicitly engage in social activities, such as group activities and making friends. Further inquiry is needed to understand several important questions, including: 1) if other existing scholarly practices (e.g. academic conferences, workshops, or electronic mailing lists) fulfill these motivations adequately without depending upon an ASNS group; 2) are there more desirable social features for academics; 3) if reluctance to engage socially is a sign of the deep-rooted nature of scholars.

Through an environmental scan about scholars engaged in an ASNS, we obtained better understanding of the user characteristics and practices on the site. Our study also illuminates that ASNS users have different or even multiple motivations for creating or joining open groups. However, the incentives to meet these different motivations have yet to be fully studied.

Future work includes the further examination of user participation and actual interactions in ASNS groups; the exploration of the reasons for the uneven distribution of academic disciplines; the identification of the features that academic users prefer and the incentives that motivate them to further engage with ASNSs; and finally, the roles that librarians and information professionals will play in ASNS groups, which may have a huge impact on the future research and development of academic social networking services.

**REFERENCES**

Bar-Ilan, J., Haustein, S., Peters, I., Priem, J., Shema, H. & Terliesner, J. (2012). Beyond citations: scholars' visibility on the social web. In Proceedings of the 17th International Conference on Science and Technology Indicators. Montréal: Science-Metrix and OST.

Bateman, P., Gray, P., & Butler, B. (2006). Community commitment: How affect, obligation, and necessity drive online behaviors. International Conference on Information Systems ICIS, 2006, 983–1000.

Borgman, C. L. (2007). Scholarship in the digital age: Information, infrastructure, and the Internet. Issues in Science Technology Librarianship (Vol. 57, p. 336). MIT Press.

Borrego, A. & Fry, J. (2012). Measuring researchers' use of scholarly information through social bookmarking data: A case study of BibSonomy. Journal of Information Science, 38(3), 297-308.

Bullinger, A. C., Hallerstede, S. H., Renken, U., Soeldner, J.-H., & Moeslein, K. M. (2010). Towards research collaboration – a taxonomy of social research network sites. Proceedings of the 16th Americas Conference on Information Systems (AMCIS). Lima, Peru.

Butler, B., Sproull, L., Kiesler, S., & Kraut, R. (2002). Community effort in online groups: Who does the work and why? (S. Weisband & L. Atwater, Eds.) Leadership at a Distance, 54(2), 346–362.

Case, D.O. (2006). Looking for information: a survey of research on information seeking, needs and behavior, 2nd ed. New York: Academic Press.

Cummings, J., Sproull, L., & Kiesler, S. (2002). Beyond hearing: Where real world and online support meet. Group Dynamics Theory Research And Practice, 6(1), 78–88.

Eysenbach, G. (2011). Can tweets predict citations? Metrics of social impact based on Twitter and correlation with traditional metrics of scientific impact. Journal of medical Internet research, 13(4), e123.

Gruzd, A. (2012). Non-academic and academic social networking sites for online scholarly communication. In D. R. Neal (Ed.), Social Media for Academics : A Practical guide. Oxford : Chandos Pub.

Hinds, P. J., & Pfeffer, J. (2003). Why organizations don't "know what they know": Cognitive and Motivational Factors Affecting the Transfer of Expertise. In M. S. Ackerman, V. Pipek, & V. Wulf (Eds.), Sharing expertise beyond knowledge management (pp. 3–26). The MIT Press.

Jeng, W., He, D., Jiang, J., & Zhang, Y. (2012b). Groups in Mendeley: Owners' descriptions and group outcomes. In ASIST 2012 Annual Meeting. Baltimore, MD.

Kanfer, R., Chen, G., & Pritchard, R. (2008). Work motivation: Forging new perspectives and directions in the post-millennium. In R. Kanfer, G. Chen, & R. D. Pritchard (Eds.), Work motivation: Past, present, and future. Routledge: New York.

Kelly, B. and Delasalle, J. (2012) Can LinkedIn and Academia.edu Enhance Access to Open Repositories? In: OR2012: the 7th International Conference on Open Repositories, Edinburgh, Scotland.

Kietzmann, J. H., Hermkens, K., McCarthy, I. P., & Silvestre, B. S. (2011). Social media? Get serious! Understanding the functional building blocks of social media. Business Horizons, 54(3), 241–251.

Kogan, M. (2000). Higher education communities and academic identity. Higher Education Quarterly, 54(3), 207–216.

Krause, J. (2012). Tracking reference with social media tools: Orgnizing what you've read or want to read. In D. R. Neal (Ed.), Social media for academics : a practical guide. Oxford : Chandos Pub.

Laird, T. F. N., Shoup, R., Kuh, G. D., & Schwarz, M. J. (2008). The effects of discipline on deep approaches to student learning and college outcomes. Research in Higher Education, 49(6), 469-494.

Lampe, C., Wash, R., Velasquez, A., & Ozkaya, E. (2010). Motivations to participate in online communities. In Proceedings of the SIGCHI Conference on Human Factors in Computing Systems. ACM.

Lampel, J., & Bhalla, A. (2007). The role of status seeking in online communities: Giving the gift of experience. Journal of Computer-Mediated Communication, 12(2), 434-455.

Latham, G. P., & Pinder, C. C. (2005). Work motivation theory and research at the dawn of the twenty-first century. Annual Review of Psychology, 56(1), 485–516.

Leahey, E. (2007). Not by productivity alone: How visibility and specialization contribute to academic earnings. American sociological review, 72(4), 533-561.

Letierce, J., Passant, A., Breslin, J., & Decker, S. (2010). Understanding how Twitter is used to widely spread Scientific Messages. In In Proceedings of the WebSci10 Extending the Frontiers of Society OnLine.

Li, X., Thelwall, M., & Giustini, D. (2012). Validating online reference managers for scholarly impact measurement. Scientometrics, 91(2), 461-471.

Meho, L. I., & Tibbo, H. R. (2003). Modeling the information-seeking behavior of social scientists: Ellis's study revisited. Journal of the American Society for Information Science and Technology, 54(6), 570–587.

Oh, J. & Jeng, W. (2011). Groups in academic social networking services - An exploration of their potential as a platform for multi-disciplinary collaboration. In IEEE SocialCom 2011. Boston, MA.

Palmer, J. (1991). Scientists and information: I. Using cluster analysis to identify information style. Journal of Documentation, 47(2), 105–129.

Porter, C. E. (2004). A typology of virtual communities: A multi-disciplinary foundation for future research. Journal of Computer-Mediated Communication, 10(1).

Priem, J., & Costello, K. L. (2010). How and why scholars cite on Twitter. Proceedings of the American Society for Information Science and Technology, 47(1), 1-4.

Priem, J., Piwowar, H. A., & Hemminger, B. H. (n.d.). Altmetrics in the wild: An exploratory study of impact metrics based on social media. Metrics 2011: Symposium on Informetric and Scientometric Research. New Orleans, LA, USA.

Schleyer, T., Spallek, H., Butler, B. S., Subramanian, S., Weiss, D., Poythress, M. L., Rattanathikun, P., & Mueller, G. (2008). Facebook for scientists: requirements and services for optimizing how scientific collaborations are established. Journal of medical Internet research, 10(3).

Shema, H., Bar-ilan, J., & Thelwall, M. (2013). Do blog citations correlate with a higher number of future citations? Research blogs as a potential source for alternative metrics.

Skeels, M. M., & Grudin, J. (2009). When social networks cross boundaries: a case study of workplace use of facebook and linkedin. Group, 10(3), 95–103.

Thelwall, M., Haustein, S., Larivière, V., & Sugimoto, C. R. (2013). Do Altmetrics work? Twitter and ten other social web services. PloS one, 8(5), e64841.

Thij, E. T. (2007). Online communities: Exploring classification approaches using participants' perspectives. In Proceedings of the IADIS International Conference on Web Based Communities.

Veletsianos, G., & Kimmons, R. (2011). Networked Participatory Scholarship: Emergent techno-cultural pressures toward open and digital scholarship in online networks. Computers & Education, 58(2), 766–774.

Weber, G. M., Barnett, W., Conlon, M., Eichmann, D., Kibbe, W., Falk-Krzesinski, H., ... & Kahlon, M. (2011). Direct2Experts: a pilot national network to demonstrate interoperability among research-networking platforms. Journal of the American Medical Informatics Association, 18(1), i157-i160.

Weller, K., Dornstädter, R., Freimanis, R., Klein, R. N., & Perez, M. (2010). Social software in academia: Three studies on users' acceptance of Web 2.0 Services. In Web Science Conference.

Weller, K., & Puschmann, C. (2011, June). Twitter for scientific communication: How can citations/references be identified and measured. In Web Science Conference.

Zaugg, H., West, R. E., Tateishi, I., & Randall, D. L. (2011). Mendeley: Creating communities of scholarly inquiry through research collaboration. TechTrends, 55(1), 32-36.

**Table A1. Sampled Open Groups by the Number of Group Members (N=97)**

| Id | Group names | No. of member |
|---|---|---|
| 574761 | Biology Classics | 284 |
| 537411 | Qualitative Research Methodology | 253 |
| 509181 | Machine Learning Basics | 246 |
| 501601 | Data analysis | 187 |
| 530031 | Future of Science | 176 |
| 536621 | Creatively named research papers | 169 |
| 602661 | Writing | 153 |
| 519291 | Social Networks - Facebook | 139 |
| 485181 | Theories of learning and teaching with technology - I | 137 |
| 492511 | Social Networks | 135 |
| 636721 | Social Network Theory: An Anthropological View | 135 |
| 595471 | Sustainability | 132 |
| 510991 | User Experience | 126 |
| 509491 | Bioinformatics | 125 |
| 485291 | Synthetic Biology | 125 |
| 630791 | Papers in the press | 122 |
| 514181 | Medical Biostatistic | 108 |
| 600591 | Computational Neuro Cognitive Science | 107 |
| 517931 | Bayesian MCMC | 100 |
| 486591 | Processes in Social Networks | 96 |
| 508851 | Knowledge Management | 96 |
| 485741 | Population Genetics | 92 |
| 493301 | Bioinformatics | 89 |
| 509151 | Strategy Research | 86 |
| 501641 | Reputation, identity and trust | 85 |
| 489051 | Games, Virtual Worlds, and CyberSociality | 77 |
| 509141 | Entrepreneurship Business Models | 76 |
| 586111 | Solar Cells, Physics and Characterization | 75 |
| 499821 | Computer Vision | 74 |
| 517191 | Behavioural Economics | 72 |
| 586171 | alt-metrics | 72 |
| 486281 | Computer Graphics | 71 |
| 524281 | Education 3.0 | 69 |
| 507531 | Twitter and Microblogging Papers | 68 |
| 531361 | How to..Biostatistics | 67 |
| 622361 | Organic User Interfaces | 65 |
| 536741 | Ecosystem Services | 64 |
| 643771 | Paleolithic Diet Research | 62 |
| 574801 | Open Access Week | 60 |
| 508971 | genomics | 60 |
| 632371 | Scientific Paper Recommender Systems | 60 |
| 516281 | Personal Learning Environments/Mashups for Learning | 57 |
| 489971 | Computational Chemistry | 55 |
| 507621 | Genome Sequencing | 52 |
| 531511 | Sentiment Analysis and Opinion Mining | 52 |
| 568281 | Physics Nobel Prizes | 52 |
| 681511 | Stockholm Resilience Centre Papers | 51 |
| 667871 | Resilience Thinking | 50 |
| 528411 | Semantic Web basics | 49 |
| 516631 | EEG neurofeedback | 49 |
| 487921 | Virtual & Augmented Reality | 46 |
| 514471 | Toxicology | 46 |
| 515011 | Systems Biology | 46 |
| 611071 | Strategy and Business | 46 |
| 484121 | Engineering Education | 45 |
| 495021 | Transcription Factor Binding Sites | 45 |
| 499111 | Business Model Innovation | 41 |
| 485041 | Numerical linear algebra and nonlinear optimization | 41 |
| 498071 | Reference Management | 37 |
| 586141 | Biomolecular Archaeology | 37 |
| 502771 | Economics | 36 |
| 485001 | Fundamentals of quantum mechanics | 36 |
| 538151 | Classics in Probability Theory | 35 |
| 528421 | Actor-Network Theory | 35 |
| 572461 | Research Methodology | 34 |
| 531641 | Research Methodology | 34 |
| 639821 | Philosophy and Education | 34 |
| 487001 | Open Innovation | 34 |
| 641411 | Cognitive Neuroscience | 33 |
| 660491 | Graduate Teaching and Mentorship for Sustainability Science | 32 |
| 510911 | Corporate Culture | 32 |
| 613231 | MSLS 2011 | 32 |
| 582941 | Finance and Corporate Finance | 31 |
| 520771 | Economics and Finance | 31 |
| 492651 | General Chemoinformatics | 30 |
| 710251 | BCB | 30 |
| 604101 | Sound and Music Computing | 30 |
| 647271 | Corporate Social Responsibility (Business Management Perspective) | 30 |
| 677641 | The Beatles | 29 |
| 483781 | Optimization | 28 |
| 585551 | Neurorobotics | 28 |
| 712841 | Library and Information Science | 28 |
| 675221 | Learning Environments | 24 |
| 510031 | Web 2.0 - Social media | 22 |
| 499531 | Robotics | 22 |
| 490481 | Digital Urbanisms - Urban Computing | 22 |
| 539491 | Corporate Entrepreneurship | 21 |
| 1066691 | Social Search | 21 |
| 549931 | Complexity | 21 |
| 524601 | Science 2.0 | 21 |
| 690261 | Eye movements and attention | 21 |
| 556371 | Innovation & Networks | 20 |
| 691941 | pharmaceutical chemistry | 20 |
| 487941 | HCI | 20 |
| 589861 | Library and Information Science | 20 |
| 507351 | Land Change Science - Public | 20 |
| 618521 | Immune-pineal axis | 20 |

Note: Data were collected in 2011 May. The URL of an open group on Mendeley is its ID followed by an address of "http://www.mendeley.com/groups/". The group Synthetic Biology with ID 485291, for example, its public URL is "http://www.mendeley.com/groups/485291/".

# APPENDIX 2

**Table A2. Survey items.**

| Respond to | Themes | Measures | ID | Items |
|---|---|---|---|---|
| **RQ1** | **Basic information** | *Discipline* | B.1.1 | User's main discipline on the research site |
| | | | B.1.2 | User's secondary discipline on the research site (dropped) |
| | | *Position* | B.2 | The user's academic position |
| | | *Gender* | B.3 | Gender |
| | | *Age* | B.4 | Birth year |
| **RQ2** | **The Extent of Use** | *Account activities* | S.1.1 | How often the user accesses their accounts |
| | | | S.1.2 | How often the user checks news feeds |
| | | | S.1.3 | How often the user updates the profiles |
| | | | S.1.4 | How long has the user been using Mendeley |
| | | | S.1.5 | How many contacts the user has |
| | **Common Ways to Use Mendeley** | *As a doc management tool* | C.1 | Organize document files such as PDFs |
| | | *As a reference manager* | C.2 | Organize/generate paper citations (i.e. reference manager) |
| | | *As a scholar search engine* | C.3 | Search for papers on Mendeley |
| | | *As an online portfolio* | C.4 | Update personal information (i.e. as your online portfolio) |
| | | *To manage existed friends* | C.5 | Keep in touch with friends in academic context |
| | | *To make more connections* | C.6 | Make more connections (e.g. meet new people in your domain) |
| **RQ3** | **The Extent of Group Use** | *Engagement of group activities* | G.1.1 | The number of *private* groups the user creates |
| | | | G.1.2 | The number of *public* groups the user creates |
| | | | G.1.3 | Being a *member* of N groups |
| | | | G.1.4 | Being a *follower* of N groups |
| | **Motivations of joining an online group** | *Information* | M1.1 | Keep up with a user's research domain |
| | | | M1.2 | Get research related questions answered |
| | | | M1.3 | Follow topics that community is paying attention to |
| | | *Networking* | M2.1 | Connect to people who have similar research interests |
| | | | M2.2 | Expand the current social network |
| | | | M2.3 | Meet more academic people |
| | | | M2.4 | Keep in touch with people one already knew |
| | | *Visibility* | M3.1 | Gain professional visibility |
| | | | M3.2 | Be present in current discussions |
| | | *Altruistic* | M4 | Contribute the reading list |

Note: Except for items B.1-B.4, S.1-S.4, we used a 5-point Likert scale to present the extent of the measured items (i.e. 1= lowest degree, 5= highest degree). This survey also contained an open-ended question for users who need to specify their own answers other than the questionnaire. We dropped the question about users' secondary discipline when analyzing the data.